\documentstyle[12pt,epsf,epsfig]{article}
\begin{document}
\begin{center}

{\bf GLUON SQUEEZED STATES IN QCD JET}

\vspace{20pt}
{\small V.KUVSHINOV, R.SHAPOROV      \\

{\it Institute of Physics, National Academy of Sciences of Belarus       \\
220072, Belarus, Minsk\\E-mail: kuvshino@dragon.bas-net.by\\
shaporov@dragon.bas-net.by}}\\

\vspace{20pt}
\end{center}
\noindent

 We study evolution of colour gluons and prove
the possibility of gluon squeezed states at the
nonperturbative QCD jet stage. Angular and rapidity
dependences of squeezed gluon second correlation function are
studied. We demonstrate that the new gluon
states can have both sub-poissonian and super-poissonian statistics
corresponding to antibunching and bunching of gluons.

\vspace{18pt}

\vspace{18pt}
{\bf 1.\quad Introduction}
\vspace{10pt}

 Analogies between multiple hadron
and photon production in quantum optics (QO) discussed long ago \cite{sv_1},
\cite{sv_2}.

In particular, squeezed states (SS), introduced by Stoler \cite{sv_stoler}
and named by Hollenhorst \cite{sv_hol}, provoke great interest.
These states can have reduced uncertainties compared with
coherent ones, sub-poissonian (for coincide phases) and
super-poissonian (for antiphases) statistics corresponding to antibunching
and bunching of photons, can decrease quantum noise \cite{sv_milb}.
The squeezed light is generated from coherent one by nonlinear
devices and is pure quantum nonperturbative phenomenon \cite{sv_milb} --
\cite{sv_kil}.

In particle physics the study of SS was stimulated by sub-poissonian
multiplicity distributions (MD)
for low energy lepton-hadron, $e^+\!e^- , p\overline p- $ collisions.
 There were also a number of phenomenological attempts to describe oscillatory
behaviour of hadron MD for different high-energy processes by general
squeezed state MD by analogy with QO \cite{sv_9} --\cite{sv_drem} .
Weakness of these approaches was in its isolation from QCD.

Studying correlations in subsystems of hadrons in different
high-energy processes led to the conclusion that for
the MD description
we should take into account of both QCD perturbative stage which
gives wide distribution with super-poissonian MD
\cite{sv_15} and nonperturbative stage, which must have sub-poissonian
MD \cite{sv_kuv} that is typical for the SS
distribution.

Quark and gluon jets in $e^+\!e^- ,$
hadron-hadron, ep scattering processes give good possibility for experimental
tests of both perturbative and nonperturbative QCD.

 The necessity of taking into account of nonperturbative QCD stage was
displayed also particularly in local parton-hadron duality, fragmentation part of
Monte-Carlo generators, power corrections, instanton contributions.

 QCD jet perturbative evolution prepares some field of gluons
\cite{sv_doksh} which then selfinteracts nonperturbatively in consequence of nonlinearities
of Hamiltonian.
Our hint is that nonperturbative selfinteraction during
jet evolution can be a source of gluon SS by
analogy with nonlinear devices in QED for photon SS.

 In this paper we study colour evolution of gluon states
at the nonperturbative stage of QCD
for the jet ring.
 We check the fulfilment of the condition of squeezing for evolved
gluon state and study the dependences of the
second correlation function on jet cone angle and rapidity.
 Some preliminary results were discussed in \cite{sv_nasa},\cite{sv_poland}.

\vspace{18pt}
{\bf 2.\quad The nonperturbative evolution of gluons}
\vspace{10pt}

 Let us consider QCD gluon jet.
 At the end of perturbative QCD cascade the factorial moments $F_q$ in jet
are close to those of a negative binomial distribution \cite{sv_web}.
Therefore gluon multiplicity
distribution is like to negative binomial one
which corresponds to specific superposition of coherent states with poissonian
multiplicity distribution \cite{sv_last}.
Since at this moment there are gluons with different colours and vector
components then at initial time there is superposition of products of
gluon coherent states
with different the colour index b and the vector component $l$ of the next
form
$\prod\limits_{b=1}^8\prod\limits_{l=1}^3\mid \alpha^b_l(0)>. $

Let us consider at first the time evolution of gluon coherent state
$\mid\alpha^b_l(0)>$ defined by the
$\mbox{Schr}\ddot{\mbox{o}}\mbox{dinger}$ equation with the
Hamiltonian~$\hat H_g $ which has the standard QCD form
\cite{sv_nasa},\cite{sv_huang}
\begin{eqnarray}
\hat H_g=\hat H_0 + \hat V&\!\!=&\!\!
\int\!\biggl\{
\frac12 \Bigl( \hat{\vec E_a} \hat{\vec E_a} +
\hat{\vec B_a} \hat{\vec B_a}\Bigr) - g\,\hat{\vec E_a}\,
C_{abc}\,\hat{\vec A_b}\hat{A_c^0} + \frac{g}2 \,\hat{\vec B_a}\,
C_{abc}
\Bigl[\hat{\vec A_b} \hat{\vec A_c}\Bigr]
\nonumber\\
&&{}+ \frac{g^2}2
\left(C_{abc}\,\hat{\vec A_b}\,\hat A_c^0\right)^2
+ \frac{g^2}2
\left(\frac12 C_{abc}
\Bigl[\hat{\vec A_b} \hat{\vec A_c}\Bigr]\right)^2
\biggr\}\, d^3\! x,
\end{eqnarray}
where $\hat H_{0} =\displaystyle
\frac12\int\!\biggl\{
\Bigl( \hat{\vec E_a} \hat{\vec E_a} +
\hat{\vec B_a} \hat{\vec B_a}\Bigr)
\biggr\}\, d^3\! x \:$ is the Hamiltonian of the "free" gluons,
$ \hat{\vec E_a} = -\vec\nabla\hat A_a^0 - \displaystyle
\frac{\partial\hat{\vec A_a}}{\partial t}, $
$\hat{\vec B_a} = \Bigl[\vec\nabla\hat{\vec A_a}\Bigr], $
$\hat{\vec A_a} $- vector potential of the gluon field,
$C_{abc}$
are structure constants of the $SU_c(3)$ group.

Let the X-axis coincides with the jet axis and the origin
of coordinates is at the beginning of the perturbative cascade.
 Then gluon momentum has next spherical coordinates:
$
\vec k=(|\vec k|\cos\theta,|\vec k|\sin\theta\,\sin\varphi,
|\vec k|\sin\theta\,\cos\varphi),
$
\quad where $0\le\theta\le\theta_{\max}$ is the angle between $\vec k$ and
jet axis, $\theta_{\max}$ is a half of the jet cone angle;
$\varphi$~is the azimuth angle ($0\le\varphi\le 2\pi$).
Assume for simplicity that
the gluons at the end of the perturbative stage of jet evolution have
close energies.
This assumption does not change the results.

Then it is easy to find that
the Hamiltonian of gluon selfinteraction
for the jet ring with cone angle $\theta $ in momentum representation
(with taking into account of the coulomb gauge condition
\footnote{Gluon squeezing is due to the Hamiltonian
nonliniarities and does not depend on gauge fixing.})
 is equal
\begin{eqnarray}
\label{v1}
\hat V&=&\frac{k_0^4}{4(2\pi)^3}\left (1-\frac{q_0^2}{k_0^2}\right )^{3/2}
g^2 \pi C_{abc}C_{adf}\Biggl\{\left (2-
\frac{q_0^2}{k^{2}_{0}}\right )\left [
a^{bcdf}_{1212}+a^{bcdf}_{1313}\right ]+a^{bcdf}_{2323}+
\nonumber\\
&&+{}\frac{\sin^2\theta}2\left (1-\frac {q_0^2}{k^{2}_{0}}
\right )
\left[2a^{bcdf}_{2323}\, -\, a^{bcdf}_{1212}\, -\, a^{bcdf}_{1313}\right]
\Biggr\}\sin\theta.
\end{eqnarray}
Here $ a^{bcdf}_{ijkj}=\hat a^{b+}_{i}\hat a^{c+}_{j}
\hat a^{d}_{k}\hat a^{f}_{j}+
\hat a^{b+}_i\hat a^c_j\hat a^{d+}_k\hat a^f_j+
\hat a^b_i\hat a^{c+}_j\hat a^{d+}_k\hat a^f_j+ c.c. ,$
$\hat a^b_i\left (\hat a^{b+}_i\right )$\, are annihilation (production)
operators of gluons,
$k_0$ and $q_0$\, are gluon energy and virtuality at the end of
perturbative cascade.
Integration over $\theta$ gives the total jet cone Hamiltonian.

The solution of the $\mbox{Schr}\ddot{\mbox{o}}\mbox{dinger}$ evolution
equation for small time has the evident form
\begin{eqnarray}
\mid\alpha^b_l (t)>&\simeq&\mid\alpha^b_l(0)> - \;
i\hat H_g \mid\alpha^b_l(0)>t
\end{eqnarray}
It gives possibility to study colour evolution of the gluon field
within a short time.
As an example consider the evolution of the gluon coherent state
with the colour index b=1 and the vector component $l=1$
\begin{eqnarray}
&&\mid\alpha^1_1 (t)>\simeq\biggl\{ 1\, -\, 2it \pi\sin\theta
\Bigl(u_3\, +\,
u_4\,\mid\alpha^1_1\mid^2\Bigr)
\biggr\}\mid\alpha^1_1 (0)>
\nonumber\\
&&{}-\, 2it \pi\sin\theta\, u_4\:\alpha^1_1\hat D\bigl(\alpha^1_1\bigr)
\mid a_1^1 (0)>
\nonumber\\
&&-\,2it \pi\, u_2
(1+u_1)\sin\theta\,\left(\alpha^1_1\right)^2
\mathop{{\sum}'}_{k=2}^7
\left(\mid\alpha^1_1 (0),2a^k_2>+\mid\alpha^1_1 (0),2a^k_3>\right)+\nonumber\\
&&{}+it \pi\, u_2 u_1
\sin^3\theta\, \left(\alpha^1_1\right)^{\!\!2}
\mathop{{\sum}'}_{k=2}^7
\Bigl(
\mid\alpha^1_1 (0),2a^k_2>\, +\,
\mid\alpha^1_1 (0),2a^k_3>\Bigr),
\end{eqnarray}
where $\mid a^b_l(0)>$ is a single gluon vector,
$\hat D(\alpha)=exp\{\alpha \hat a^+ - \alpha^*\hat a\}$ is the
displacement operator of amplitude $\alpha$,
the explicit forms of the constants $u_1,\, u_2,\, u_3,\, u_4 $ and
$ \mathop{{\sum}'}\limits_{k=2}^7\bigl( \bigr)$
are given in the Appendix.

 Analogously we can also investigate the evolution of coherent gluon states with
any other colour charges and vector components.

As the result the following conclusion has been obtained:\\
 1) for the initial vectors with the colour indexes b=1,2,3 the vectors
with another colour indexes k$=\overline{4,7}$ appear;
 2) if the initial vectors have the colour indexes b=$\overline{4,7}$
then the new vectors
with colour indexes k=1,2,3,8 and the vectors with the combination
of the colour indexes 3,8: $\mid\alpha^{b}_{l},a^{3},a^{8}>$ appear;
 3) as the result of the evolution of colour coherent state with b=8
the mixed colour states with colour indexes 4,5,6,7 appear.

It is clear that namely the difference among the structure constants of the
$SU_c(3)$-group for different colour indexes leads to
the different evolution of the corresponding colours.

\vspace{18pt}
{\bf 3.\quad Gluon squeezed state}

\vspace{10pt}
To prove that the evolved state $\mid f>$ is the gluon squeezed state (GSS)
we must check by analogy with QO the
fulfilment of the squeezing condition which has in particular the form
\cite{sv_kil}
\begin{eqnarray}
\label{form_2}
\Bigl\langle N\left({\scriptstyle\triangle}(\hat X^b_l)_{1\atop 2}\right)^2
\Bigr\rangle
=\Bigl\langle\left( {\scriptstyle\triangle}(\hat X^b_l)_{1\atop 2}\right)^2
\Bigr\rangle-\frac14\, <\, 0
\end{eqnarray}
where $\biggl\langle\left({\scriptstyle\triangle}(\hat X^b_l)_{1\atop 2}\right)^2
\biggr\rangle =
\biggl\langle\biggl((\hat X^b_l)_{1\atop 2} -
\Bigl\langle(\hat X^b_l)_{1\atop 2}\Bigr\rangle\biggr)^2 \biggr\rangle,$
averaging is made over $\mid f>,$
real and imaginary components of the complex amplitude of the gluon field
are defined by the operators
$(\hat X^b_l)_1=\big[\hat a^b_l\, +\, (\hat a^b_l)^+\big] /2 $ and
$(\hat X^b_l)_2=$\\
$=\big[\hat a^b_l\, -\, (\hat a^b_l)^+\big] /2i, $
the operator of normal ordering N is
\begin{eqnarray}
\label{form_3}
\Bigl\langle N\left({\scriptstyle\triangle}(\hat X^b_l)_{1\atop 2}\right)^2
\Bigr\rangle &=&
\frac14
\Biggl\{\,\pm\,\left[\Bigl\langle\Bigl(\hat a^b_l\Bigr)^2\Bigr\rangle
-\Bigl\langle\hat a^b_l\Bigr\rangle^2\right]\pm
\left[\Bigl\langle\Bigl(\hat a^{b+}_l\Bigr)^2\Bigr\rangle
-\Bigl\langle\hat a^{b+}_l\Bigr\rangle^2\right]\nonumber\\
&&{}+ 2\left[\Bigl\langle\hat a^{b+}_l\hat a^b_l\Bigr\rangle\, -
\Bigl\langle\hat a^{b+}_l\Bigr\rangle
\Bigl\langle\hat a^b_l\Bigr\rangle\right]
\Biggr\}.
\end{eqnarray}
  To check the squeezing of the final evolved state
it is sufficient to make averaging of the
$N\left({\scriptstyle\triangle}(\hat X^b_l)_{1\atop 2}\right)^2 $
over the vector
$\prod\limits_{b=1}^8\prod\limits_{l=1}^3\mid \alpha^b_l(t)>. $
In this case it is easy to see that
$\Bigl\langle N\left({\scriptstyle\triangle}(\hat X^b_l)_{1\atop 2}\right)^2
\Bigr\rangle\ne 0.$
From the explicit form of
$\Bigl\langle N\left({\scriptstyle\triangle}(\hat X^b_l)_{1\atop 2}\right)^2
\Bigr\rangle $
for the colour index b=1 and an arbitrary vector
component~$l$
\begin{eqnarray}
\Bigl\langle N\left({\scriptstyle\triangle}(\hat X_l^1)_{1\atop 2}\right)^2
\Bigr\rangle&=&
\pm \: 4\pi\, u_2\, t\, d\theta
\biggl\{
(1+u_1)
\sin\theta
\Bigl[
\delta_{l1}(Z_{33}+Z_{22})+(1-\delta_{l1})Z_{11}
\Bigr]
\nonumber\\
&&+(1-\delta_{l1})\sin\theta\Bigl[\delta_{l2}
Z_{33}+\delta_{l3}Z_{22}\Bigr]
+u_1\sin^3\theta\:\Bigl[-\frac12\delta_{l1}\times
\nonumber\\
&&(Z_{22}+Z_{33})+\delta_{l2}\:(Z_{33}-\frac12 Z_{11})+\delta_{l3}
\:(Z_{22}-\frac12 Z_{11}
)\Bigr]
\biggr\},
\end{eqnarray}
$(Z_{mn}=\mathop{{\sum}'}\limits_{k=2}^7\bigl\langle (\hat X_m^k)_1\bigr\rangle
\bigl\langle (\hat X_n^k)_2\bigr\rangle,\: m,n=1,2,3) $
one can see that the squeezing condition (\ref{form_2}) is
fulfiled with uncertainties
${\scriptstyle\triangle}(\hat X^1_l)_2\, <\,\displaystyle\frac14\,
<{\scriptstyle\triangle}
(\hat X^1_l)_1 $ under the next conditions:
 $\bigl\langle (\hat X^1_l)_1\bigr\rangle<0 $,
$\bigl\langle (\hat X^1_1)_2\bigr\rangle<0 $ or
 $ \bigl\langle (\hat X^1_l)_1\bigr\rangle>0 $,
$ \bigl\langle (\hat X^1_l)_2\bigr\rangle>0 $  and
with uncertainties
${\scriptstyle\triangle}(\hat X^1_l)_1\, <\,\displaystyle\frac14\,
<{\scriptstyle\triangle}(\hat X^1_l)_2 $
under the next conditions:\\
$\bigl\langle (\hat X^1_l)_1\bigr\rangle>0 $,
$\bigl\langle (\hat X^1_l)_2\bigr\rangle<0 $ or
 $\bigl\langle (\hat X^1_l)_1\bigr\rangle<0 $,
$\bigl\langle (\hat X^1_l)_2\bigr\rangle>0 .$

Thus the evolved vector
$\prod\limits_{b=1}^8\prod\limits_{l=1}^3\mid \alpha^b_l(t)> $
or its combinations can describe the GSS.

Note that if we make averaging over the superposition of the
gluon coherent states
with fixed colour and vector components,
or the superposition
of the gluon coherent states with fixed colour,
or fixed vector component,
or superposition of the single gluon states,
then we can see that the quantity
$ \Bigl\langle N\left({\scriptstyle\triangle}(\hat X^b_l)_{1\atop 2}\right)^2
\Bigr\rangle $ is equal zero, the condition (\ref{form_2}) is not
fulfiled and these states are not GSS.

\vspace{18pt}
{\bf 4.\quad
Two gluon correlation function
}

\vspace{10pt}

 What could be an experimental indication on GSS ? To answer this question,
we can study the angular dependence of squeezed gluon second correlation
function by well-known methods of QO.

 By analogy with QO we can write the second normalized correlation function
of gluons
in the form
\begin{eqnarray}
K^{b}_{l(2)} (\theta_1,\theta_2) &=&\frac{\left\langle \hat a^{b+}_l\,\hat a^{b+}_l\,\hat a^b_l\,
\hat a^b_l\right\rangle}{\left\langle \hat a^{b+}_l\,\hat a^b_l\right\rangle^2}
- 1
\end{eqnarray}
The averaging here as it must is carried out over the state vector\\
$\prod\limits_{b=1}^8\prod\limits_{l=1}^3\mid\! \alpha ^b_l(\theta_1,t),
\alpha ^b_l(\theta_2,t)\!> $
at the moment t.
If $K^{b}_{l(2)} > 0 $ then bunching of gluons takes
place and the gluon antibunching can occur in the case $K^{b}_{l(2)} < 0. $
For a coherent field with a poissonian distribution of
gluons $K^{b}_{l(2)} $ is equal 0.
 At the beginning of the nonperturbative region
$K^{b}_{l(2)} = 0$
because the gluon state vector at the initial moment is the product of the
gluon coherent states.

Averaging over the evolved vector
$\prod\limits_{b=1}^8\prod\limits_{l=1}^3\mid\! \alpha ^b_l(\theta_1,t),
\alpha ^b_l(\theta_2,t)\!> $
which also describes gluon squeezed state, we obtain
\begin{eqnarray}
\label{form_7}
K^{b}_{l(2)}(\theta_1,\theta_2) &=&-\,\frac{M_1 (\theta_1,\theta_2)}
{\mid\alpha^b_l\mid^4-2\mid\alpha^b_l\mid^2M_1 (\theta_1,\theta_2)+
M_2 (\theta_1,\theta_2)},
\end{eqnarray}
where for the colour
1 and an arbitrary vector component~$l$
\begin{eqnarray}
M_1(\theta_1,\theta_2) = 24\,t\,u_2\:\pi\mid\alpha\mid^2\mid\beta\mid^2
\sin(2\delta+\pi /2)\Bigl\{(1+\delta_{l1})(2+u_1-\delta_{l1})
 \nonumber\\
\times
(\sin\theta_1\,+\sin\theta_2)
- \frac12\,u_1\:(3\delta_{l1} - 1) (\sin^3\theta_1\,+\sin^3\theta_2)
\Bigr\},
\end{eqnarray}
\begin{eqnarray}
M_2(\theta_1,\theta_2) = 80\,t\,u_2\:\pi\mid\alpha\mid^3\mid\beta\mid^3
\sin(\delta+\pi /4)\Bigl\{(1+\delta_{l1})(2+u_1-\delta_{l1})
 \nonumber\\
\times (\sin\theta_1\,+
\sin\theta_2)
- \frac12\,u_1\:(3\delta_{l1} - 1) (\sin^3\theta_1\,
+\sin^3\theta_2)
\Bigr\}
\end{eqnarray}
Here for symplicity we supposed
that $\alpha^1_l=\mid\alpha\mid e^{i\gamma_1} $ for $\forall \,l\:$
and $\alpha^b_l=\mid\beta\mid
e^{i\gamma_2}, $ when $b\ne 1, $ for $\forall \,l,\: \gamma_1 - \gamma_2 =
\delta + \displaystyle\frac{\pi}4$ (phase $\delta$ defines the direction of
squeezing maximum \cite{sv_milb}).

 At the same time for the squeezed state of photons
the second normalized correlation function at
$0 < r_l < \displaystyle\frac14 $ ($r_l$ is squeezing parameter for
component~$l$) is \cite{sv_milb},\cite{sv_kil}
\begin{eqnarray}
\label{form_6}
K^{b}_{l(2)} &=&-\,\frac{r_l[\alpha^2_l e^{-2i\delta}+(\alpha^*_l)^2
e^{2i\delta}]}
{\mid\alpha_l\mid^4-2r_l\mid\alpha_l\mid^2
[\alpha_l^2 e^{-2i\delta}+(\alpha^*_l)^2 e^{2i\delta}]}
\end{eqnarray}
 It can be both less than 0 in the case of phase
$\delta = 0 $ (coincide phases), and more than 0 in the case of phase
$\delta = \displaystyle\frac{\pi}2 $ (antiphases)
corresponding to antibunching
(sub-poissonian MD) and bunching (super-poissonian MD) of photons
\cite{sv_milb}.

 Unlike corresponding expression (\ref{form_6}) in QO
$K^{b}_{l(2)}(\theta_1,\theta_2) $ for GSS (\ref{form_7}) includes also
function $M_2 (\theta_1,\theta_2) $ which appears due to the
different colours and vector components of gluons in the
Hamiltonian.

We can write correlation function in the terms of the rapidity
$K^{b}_{l(2)}(y_1,y_2) $ by transformation
\begin{equation}
\begin{array}{c}
\sin\theta= \sqrt{1-\displaystyle\frac{\tanh^2 y}{u_1}},\quad
d\theta= - \displaystyle\frac{dy}{\cosh^{2}y\sqrt{u_1 - \tanh^2 y}}
\end{array}
\end{equation}

The angle and the rapidity dependences of squeezed gluon correlation function
is plotted
for b=1 at the time t=0.001, $\theta_2 = 0 $ (Fig.1) and $y_2 = 0$ (Fig.2)
under some reasonable parameters:
$g=1,$ that corresponds to the bound between perturbative
and nonperturbative regions;
$q_0^2 = 1\: GeV^2$ that correponds to the gluon virtuality at the beggining of
the nonperturbative stage;
$k_0=\displaystyle\frac{\sqrt{s}}{2<n_{gluon}>} \:$ corresponds to
a gluon energy; $\sqrt{s} = 91\: GeV$ and $<n_{gluon}> = 10$.

From the Fig.1 we notice that angle correlations have singularity at
$\theta_1\sim 10^{-9}$ and then with increasing $\theta_1 $
it decreases and approaches to the value $ 0.522.$
Besides $K^{1}_{l(2)}(\theta_1,\theta_2=0)$ has peak
$K^{1}_{l(2)}(0,0)=0$ at the beginning of the nonperturbative stage
$(\theta_1=\theta_2=0) $ because the gluon state vector at the initial
moment is the product of the gluon coherent states.

At the same parameters the rapidity correlation function has two
maximums
$K^{1}_{l(2)}(y_1,y_2=0)=-2\cdot 10^{-5}$ at
$y_1 = \pm 1.98$
and minimum\\
$K^{1}_{l(2)}(y_1=0,y_2=0)=-7.295\cdot 10^{-5}$ (Fig.2).
Since the rapidity correlations fall in the negative region then we have
antibunching of gluons at the nonperturbative stage of QCD jet.
\begin{figure}
\epsfxsize=2.5in \epsfbox{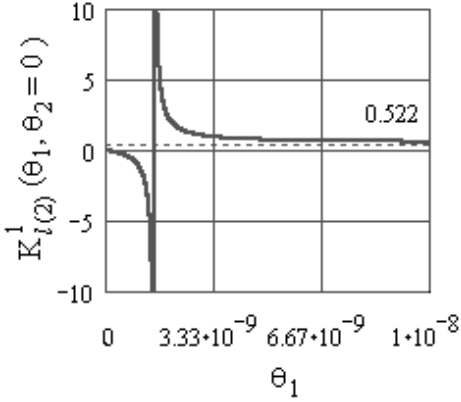}
\caption{The angular dependence of the squeezed gluon correlation function
$K^{1}_{l(2)}(\theta_1)$ at $\theta_2~=~0,K^{1}_{l(2)}~(0,0)~=~0.$  }
\end{figure}
\begin{figure}
\epsfxsize=2.5in \epsfbox{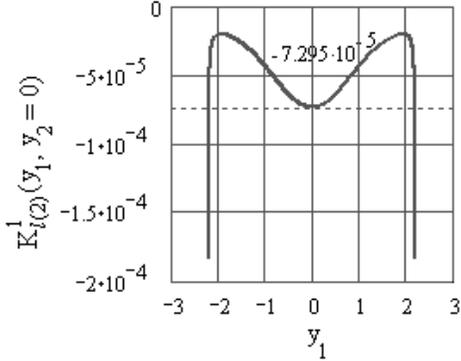}
\caption{The rapidity dependence of the squeezed gluon correlation function
$ K^{1}_{l(2)}(y_1)$ at $y_2~=~0,
K^{1}_{l(2)}~(0,0)~=0.$  }
\end{figure}

\vspace{18pt}
{\bf 5.\quad
Conclusion
}

\vspace{10pt}

Thus nonperturbative quantum evolution of gluon state prepared by perturbative
cascade stage in jets can lead at least in a small time
to quantum gluon states - squeezed states.

 The two gluon correlation function in GSS is calculated by analogy with
QO. It was
demonstrated that the second correlation function
 $K^{1}_{l(2)}(\theta_1,\theta_2~=~0~) $
has singularity at some angle and then with increasing $\theta_1 $ it
decreases and approaches to the constant value. At the same time the rapidity
correlations fall into the negative region and have minimum at $y_1 = 0. $
This behavior corresponds to antibunching of gluons this is a sub-poissonian
 multiplicity distribution at the nonperturbative stage of QCD jet.

Such behaviour of the correlation functions can be the sign
of gluon squeezed states.
This effect can be searched experimentally at LEP and other facilities
where jets are clearly seen under adequate
taking into account the perturbative cascade correlations and hadronization.

\vspace{18 pt}
\noindent
{\bf Acknowledgments}
\vspace{10pt}

\noindent
The authors are grateful for support in part to Basic Science Foundation
of Belarus (Projects F95-023, M96-023).

\vspace{18 pt}
\noindent
{\bf Appendix }
$$
\mathop{{\sum}'}\limits_{k=2}^7\bigl( \bigr) =
\sum\limits_{k=2}^3 \bigl( \bigr)+\frac14\sum\limits_{k=4}^7
\bigl( \bigr),
\;u_1 = \biggl(1-\displaystyle\frac{q_0^2}{k^2_0}\biggr),
\;u_2 =
\frac{k^4_0}{4(2\pi)^3}\frac{g^2}2
(u_1)^{\frac32},
$$
$$
u_3 = \displaystyle\frac{k^3_0}2
(u_1)^{\frac12}\,
15\biggl(1+u_1+\frac{q_0^4}{2k^4_0}\biggr)\,+
\,24\,u_2 (3+2u_1),
$$
$$
u_4 = \displaystyle\frac{k^3_0}2
(u_1)^{\frac12}\,
\Biggl\{\frac{q_0^4}{k^4_0}\,+\,
\biggl(1+u_1-\frac{q_0^4}{k^4_0}\biggr)\sin^2\theta\Biggr\}
 +\,6\,u_2[2(1+u_1)-
u_1\sin^2\theta ].
$$

\end{document}